\documentclass[11pt]{article}
\usepackage{amsmath, amsthm, amsfonts}
\usepackage[margin=1in,footskip=0.25in]{geometry}
\usepackage{makecell}
\usepackage{multicol}
\usepackage{graphicx}
\theoremstyle{theorem}

\usepackage{hyperref}
\usepackage{cite}

\newtheoremstyle{defi}
  {10pt}          
  {10pt}  
  {\rm}  
  {\parindent}     
  {\bf}  
  {. }    
  { }    
  {}     
\theoremstyle{defi}

\begin{document}

\date{}

\title{\bf Dark energy in spherically symmetric universe coupled with Brans-Dicke scalar field}
\author{Koijam Manihar Singh$^{1}$, Gauranga C. Samanta$^{2}$ \\
 $^{1}$ Department of Mathematics, Mizoram University,Aizawl, Mizoram-796004, India \\
 $^{2}$Department of Mathematics,
BITS Pilani K K Birla Goa Campus,
Goa-403726, India \\ drmanihar@rediffmail.com\\ gauranga81@gmail.com}

\maketitle

\begin{abstract}
The phenomenon of dark energy and its manifestations are studied in a spherically symmetric universe considering the Brans-Dicke
scalar tensor theory. In the first model the dark energy behaves like a phantom type and in such a universe the existence of negative time is validated with an indication that our universe started its evolution before $t=0$. Dark energy prevalent in this universe is found to be more active at times when other types of energies remain passive. The second model universe begins with  big bang. On the other hand the dark energy prevalent in the third model is found to be of the quintessence type. Here it is seen that the dark energy triggers the big bang and after that much of the dark energy reduces to dark matter. One peculiarity in such a model is that the scalar field is prevalent eternally, it never tends to zero.
\end{abstract}


\textbf{Keywords}: Dark energy, Brans-Dicke theory, Big Bang. \\
\textbf{Mathematics Subject Classification Codes:} 83C05; 83C15; 83F05.


\section{Introduction}
The type Ia (SNIa) supernovae observations suggested that our universe is not only expanding, but also the rate of expansion is in accelerating way \cite{Riess, Perlmutter} and this acceleration is caused by some mysterious object so called dark energy. The matter species in the universe are broadly classified into relativistic particle, non relativistic particle and dark energy. Another component, apparently a scalar field, dominated during the period of inflation in the early universe. In the present universe, the sum of the density parameters of baryons, radiation and dark matter does not exceed 30\% \cite{Amendola}, we still need to identify the remaining 70\% of the cosmic matter. We call this 70\% unknown component is dark energy, and it is supposed to be responsible for the present cosmic acceleration of the universe. According to the cosmological principle our universe is homogeneous and isotropic in large scale. By assuming isotropicity and homogeneity, the acceleration equation of the universe in general theory of relativity can be written as $\frac{\ddot{a}}{a}=-\frac{1}{6}\kappa^2(\rho+3p)$. The acceleration and deceleration of the universe depend on the sign of $\ddot{a}$, i. e. the universe will accelerate if $\rho+3p<0$ or decelerate if $\rho+3p>0$. So, the condition $\rho+3p<0$ has to be satisfied in general relativity to explain accelerated expansion of the universe. This implies that the strong energy condition is violated, moreover the strong energy condition is violated means, the universe contains some abnormal (something not normal) matter. Hence, without violating strong energy condition, the accelerated expansion of the universe is not possible in general theory of relativity. Therefore, the modification of the general theory of relativity is necessary. Essentially, there are two approaches, out of which one is: to modify the right hand side of the Einstein's field equations (i. e. matter part of the universe) by considering some specific forms of the energy momentum tensor $T_{\mu\nu}$ having a huge negative pressure, and which is concluded in the form of some mysterious energy dubbed as dark energy. In this approach, the simplest candidate for dark energy is cosmological constant $\Lambda$, which is described by the equation of sate $p=-\rho$ \cite{Weinberg}. The second approach is by modifying Einstein Hilbert action, i. e. the geometry of the space-time, which is named as modified gravity theory. So many modifications of general relativity theory has been done, namely Brans-Dicke (BD) \cite{Brans} and Saez-Ballester scalar-tensor theories \cite{Saez}, $f(R)$ gravity \cite{Barraco,Capozziello,Capozziello1,Capozziello2,Nojiri,Bamba}, $f(T)$ gravity
\cite{Ferraro, Myrzakulov, Bamba1, Setare}, Gauss-Bonnet theory \cite{Nojiri1, Sokolowski, Iglesias, Nojiri2}, Horava-Lifshitz gravity
\cite{Colgain, Ghodsi, Park} and recently $f(R, T)$ gravity \cite{Harko}. Subsequently, so many authors \cite{Dehghani, Nojiri3, Lue, Carroll, Samanta, Samanta1, Samanta2, Samanta3, Singh, Singh1, Bamba3, Zubair, Maharaj, Priyokumar, Samanta5, Morais, Bamba2, Noureen, Samanta4, Baffou, Azmat, Singh2, Odintsov, Addazi, Shabani, Odintsov1, Zubair1, Kleidis, Tiwari, Hansraj} have been studying in modified gravity theory to understand the nature of the dark energy and accelerated expansion of the universe.

Apart from this the Hubble parameter $H$ may provide some important information about the evolution of our universe. It is dynamically determined by the Friedmann equations, and then evolves with cosmological red-shift. The evolution of Hubble parameter is closely related with radiation, baryon,
cold dark matter, and dark energy, or even other exotic components available in the universe. Further, it may be impacted
by some interactions between these cosmic inventories. Thus, one can look out upon the evolution of the universe by studying the Hubble parameter. Besides dark energy there exists a dark matter component of the universe. One can verify whether these two components can interact with each other. Theoretically,
there is no evidence against their interaction. Basically, they may exchange their energy which affects the cosmic evolution of the universe.
Furthermore, it is not clear, whether the non-gravitational interactions between two energy sources produced by two different matters in our universe can produce acceleration. We can assume for a while that the origin of non-gravitational interaction is related to emergence of the space-time dynamics. However, this is not of much help, since this hypothesis is not more fundamental compared with other phenomenological assumptions within modern cosmology
\cite{Rowland, Cui, Baldi, Sadeghi, Khurshudyan}. However, the authors \cite{Bertolami, Olivares} studied the finding that the interacting cosmological models make good agrement with observational data. The aim of this paper is to study a cosmological model, where a phenomenological form of
non-gravitational interactions are involved. In this article we are interested in the problem of accelerated
expansion of the large scale universe, we follow the well known approximation of the energy content of the
recent universe. Namely, we consider the interaction between dark energy and other matters (including dark matter).

\section{Space-time and field equations}
We consider the spherically symmetric space time
\begin{equation}\label{}
  ds^2=dt^2-e^{\lambda}(dr^2+r^2d\theta^2+r^2\sin^2\theta d\phi^2),
\end{equation}
where $\lambda$ being a function of time. The energy momentum tensor for the fluid comprising of our universe is taken as
\begin{equation}\label{}
  T_{\mu\nu}=(p+\rho)u_{\mu}u_{\nu}-pg_{\mu\nu},
\end{equation}
where $\rho$ and $p$ are respectively the total energy density and total pressure which are taken as
\begin{equation}\label{}
  \rho=\rho_m+\rho_d
\end{equation}
and
\begin{equation}\label{}
  p=p_m+p_d,
\end{equation}
$\rho_d$ and $\rho_m$ being respectively the densities of dark energy and other matters in this universe. $p_d$ and $p_m$ are respectively the pressures of dark energy and other matters (including dark matter) in this universe. And $u_{\mu}$ is the flow vector satisfying the relations
\begin{equation}\label{}
  u_{\mu}u^{\mu}=1;~~~ u_{\mu}u^{\nu}=0.
\end{equation}
The Brans-Dicke scalar tensor field equations are given by
\begin{equation}\label{}
  R_{\mu\nu}-\frac{1}{2}Rg_{\mu\nu}=-8\pi\phi^{-1}T_{\mu\nu}-\omega\phi^{-2}(\phi_{,\mu}\phi_{,\nu}-\frac{1}{2}g_{\mu\nu}\phi_{,\gamma}\phi^{,\gamma})
  -\phi^{-1}(\phi_{\mu;\nu}-g_{\mu\nu}\phi_{;\gamma}^{,\gamma})
\end{equation}
with
\begin{equation}\label{}
  \phi_{;\gamma}^{,\gamma}=8\pi(3+2\omega)^{-1}T,
\end{equation}
where $\omega$ is the coupling constant and $\phi$ is the scalar field. Energy conservation gives the equation
\begin{equation}\label{}
  T^{\mu\nu}_{;\mu}=0
\end{equation}
Here the field equations take the form
\begin{equation}\label{9}
  \frac{3}{4}\dot{\lambda}^2-\frac{\omega}{2}\frac{\dot{\phi}^2}{\phi^2}+\frac{3}{2}\dot{\lambda}\frac{\dot{\phi}}{\phi}=
  8\pi\phi^{-1}(\rho_m+\rho_d)
\end{equation}

\begin{equation}\label{10}
  \ddot{\lambda}+\frac{3}{4}\dot{\lambda}^2+\frac{\omega}{2}\frac{\dot{\phi}^2}{\phi^2}+\dot{\lambda}\frac{\dot{\phi}}{\phi}+ \frac{\ddot{\phi}}{\phi}
  =-8\pi \phi^{-1}(p_m+p_d)
\end{equation}

\begin{equation}\label{11}
  \frac{3}{2}(\ddot{\lambda}+\dot{\lambda}^2)+\frac{\omega}{2}\frac{\dot{\phi}^2}{\phi^2}+\frac{\ddot{\phi}}{\phi}+
  \frac{3}{2}\dot{\lambda}\frac{\dot{\phi}}{\phi}=-8\pi \phi^{-1}(p_m+p_d)
\end{equation}

\begin{equation}\label{12}
  \ddot{\phi}+\frac{3}{2}\dot{\lambda}\dot{\phi}=8\pi(3+2\omega)^{-1}(\rho_m+\rho_d-3p_m-3p_d)
\end{equation}
Here we take the equation of state parameter for dark energy as $\alpha$ so that
\begin{equation}\label{13}
  p_d=\alpha\rho_d.
\end{equation}
And the conservation equation gives
\begin{equation}\label{14}
  \dot{\rho}+(p+\rho)\frac{3}{2}\dot{\lambda}=0
\end{equation}
Since the dark energy and other matters are interacting in this universe equation \eqref{14} can be written as
\begin{equation}\label{15}
  \dot{\rho_m}+(\rho_m+p_m)\frac{3}{2}\dot{\lambda}=-Q
\end{equation}
and
\begin{equation}\label{16}
  \dot{\rho_d}+(\rho_d+p_d)\frac{3}{2}\dot{\lambda}=Q,
\end{equation}
where $Q$ is the interaction between dark energy and other matters (including dark matter) which comprises of this universe. Here $Q$ can take different forms like $3z^2\rho$, $3z^2\rho_m$, $3z^2\rho_d$ etc., where $z^2$ is a coupling constant. It can also take other forms which are functions of $\rho$
 and $\dot{\rho}$. Now from equations \eqref{10} and \eqref{11} we get the relation
 \begin{equation}\label{17}
   \ddot{\lambda}+\frac{3}{4}\dot{\lambda}^2+\dot{\lambda}\frac{\dot{\phi}}{\phi}=\frac{3}{2}\ddot{\lambda}+\frac{3}{2}\dot{\lambda}^2+\frac{3}{2}
   \dot{\lambda}\frac{\dot{\phi}}{\phi}
 \end{equation}
 which gives
 \begin{equation}\label{18}
   e^{\frac{3}{2}\lambda}\dot{\lambda}=a_0\phi^{-1}
 \end{equation}
  where $a_0$ is an arbitrary constant.
  \section{Analytical solutions}
  In this section, we try to obtain the analytical solutions of the field equations in three different cases based on the different forms of the interaction
  parameter $Q$.
  \subsection{Case-I:}
  From equations \eqref{17} and \eqref{18} we get
  \begin{equation}\label{}
    \lambda=\frac{2}{3}\log b_2+\log(b_0+b_1t)^{a_1}
  \end{equation}

  \begin{equation}\label{}
    \phi=\frac{a_0}{a_1b_1b_2}(b_0+b_1t)^{1-\frac{3}{2}a_1}
  \end{equation}
  where $a_1$, $b_0$, $b_1$ and $b_2$ are arbitrary constants. Here in this case we take

  \begin{equation}\label{}
    Q=3z^2H\rho,
  \end{equation}
  where $H$ is the Hubble's parameter so that the conservation equation takes the form of the equations
  \begin{equation}\label{22}
    \dot{\rho_m}+(\rho_m+p_m)\frac{3}{2}\dot{\lambda}=-\frac{3}{2}z^2\dot{\lambda}\rho
  \end{equation}
and
  \begin{equation}\label{23}
    \dot{\rho_d}+(\rho_d+p_d)\frac{3}{2}\dot{\lambda}=\frac{3}{2}z^2\dot{\lambda}\rho.
  \end{equation}
  Now from equation \eqref{23} we get

  \begin{equation}\label{24}
    \rho_d=b_4(b_0+b_1t)^{-\frac{3}{2}(\alpha+1)a_1}+b_3(1-\frac{3}{2}a_1\alpha)^{-1}(b_0+b_1t)^{-1-\frac{3}{2}a_1}
  \end{equation}
  where $b_4$ is an arbitrary constant and
  \begin{equation}\label{}
    b_3=\frac{3z^2a_0b_1}{16b_2\pi}\bigg[\frac{4}{3}a_1^2-\frac{\omega}{2}\left(1-\frac{3}{2}a_1\right)^2
    +\frac{3}{2}a_1\left(1-\frac{3}{2}a_1\right)\bigg]
  \end{equation}
  Thus using equation \eqref{24} in equation \eqref{9} we have

  \begin{equation}\label{26}
    \rho_m=\frac{2b_3}{3a_1z^2}(b_0+b_1t)^{-1-\frac{3}{2}a_1}+b_3\left(\frac{3}{2}a_1\alpha-1\right)^{-1}(b_0+b_1t)^{-1-\frac{3}{2}a_1}
    -b_4(b_0+b_1t)^{-\frac{3}{2}(1+\alpha)a_1}
  \end{equation}
  Therefore equation \eqref{22} gives
  \begin{eqnarray} \label{27}
    p_m &=& \bigg[\frac{z^2a_0b_1}{8\pi a_1b_2}\{\frac{\omega}{2}\left(1-\frac{3}{2}a_1\right)^2-\frac{4}{3}a_1^2-\frac{3}{2}a_1\left(1-\frac{3}{2}a_1\right)\} \nonumber \\
     &-& \frac{2}{3a_1b_1}\{\frac{2b_1b_3}{3a_1z^2}+b_1b_3\left(\frac{3}{2}a_1\alpha-1\right)^{-1}\}-\frac{2b_3}{3a_1z^2} \nonumber \\
     &-& b_3\left(\frac{3}{2}a_1\alpha-1\right)^{-1}\bigg](b_0+b_1t)^{-1-\frac{3}{2}a_1}+
     \left(b_4+\frac{2b_4}{3a_1}\right)(b_0+b_1t)^{-\frac{3}{2}(1+\alpha)a_1}
  \end{eqnarray}
  Now using equation \eqref{27} in equation \eqref{10} we have

  \begin{eqnarray}\label{28}
    p_d &=& \bigg[\frac{z^2a_0b_1}{8\pi a_1b_2}\{\frac{4}{3}a_1^2-\frac{\omega}{2}\left(1-\frac{3}{2}a_1\right)^2+
    \frac{3}{2}a_1\left(1-\frac{3}{2}a_1\right)\} \nonumber \\
     &+& \frac{2}{3a_1b_1}\{\frac{2b_1b_3}{3a_1z^2}+b_1b_3\left(\frac{3}{2}a_1\alpha-1\right)^{-1}\}+
     b_3\left(\frac{3}{2}a_1\alpha-1\right)^{-1} \nonumber \\
    &+& \frac{2b_3}{3a_1z^2}-\frac{a_0}{8\pi a_1b_1b_2}\{\frac{3}{4}a_1^2b_1^2-a_1b_1^2+\frac{\omega}{2}b_1^2
    \left(1-\frac{3}{2}a_1\right)^2 \nonumber \\
    &+&a_1b_1^2\left(1-\frac{3}{2}a_1\right)-b_1^2\left(1-\frac{3}{2}a_1\right)\}\bigg](b_0+b_1t)^{-1\frac{3}{2}a_1} \nonumber \\
    &-& b_4\left(1+\frac{2}{3a_1}\right)(b_0+b_1t)^{-\frac{3}{2}(1+\alpha)a_1}
  \end{eqnarray}
  Again from \eqref{13} and \eqref{24} we get
  \begin{equation}\label{29}
    p_d=\alpha b_4(b_0+b_1t)^{-\frac{3}{2}(1+\alpha)a_1}+\alpha b_3(1-\frac{3}{2}a_1\alpha)^{-1}(b_0+b_1t)^{-1-\frac{3}{2}a_1}
  \end{equation}
  Thus comparing coefficients of $(b_0+b_1t)^{-\frac{3}{2}(1+\alpha)a_1}$ and $(b_0+b_1t)^{-\frac{3}{2}a_1-1}$ of the two expressions of $p_d$ in
  \eqref{28} and \eqref{29} we obtain

  \begin{eqnarray} \label{30}
     && \frac{z^2a_0a_1b_1}{6\pi b_2}-\frac{z^2\omega a_0b_1}{16\pi a_1b_2}(1-\frac{3}{2}a_1)^2+\frac{z^2a_0b_1}{16\pi b_2}(1-\frac{3}{2}a_1) \nonumber \\
    &+& \frac{4b_3}{9a_1^2z^2}+\frac{2b_3}{3a_1}(\frac{3}{2}a_1\alpha-1)^{-1}+\frac{2b_3}{3a_1z^2}+b_3(\frac{3}{2}a_1\alpha-1)^{-1} \nonumber \\
    &-& \frac{3a_0a_1b_1}{32\pi b_2}+\frac{a_0b_1}{8\pi b_2}-\frac{\omega a_0b_1}{16\pi a_1b_2} (1-\frac{3}{2}a_1)^2 \nonumber \\
     &-& \frac{a_0b_1}{8\pi b_2}(1-\frac{3}{2}a_1)+\frac{a_0b_1}{8\pi a_1b_2}(1-\frac{3}{2}a_1) \nonumber \\
     &=& \frac{z^2a_0a_1b_1}{6\pi b_2}-\frac{z^2\omega a_0b_1}{16\pi a_1b_2}(1-\frac{3}{2}a_1)^2+ \frac{z^2 a_0b_1}{16\pi b_2}(1-\frac{3}{2}a_1)
     \nonumber\\
    &+& \frac{4b_3}{9a_1^2z^2}+\frac{2b_3}{3a_1z^2}+b_3(\frac{3}{2}a_1\alpha-1)^{-1}(\frac{2}{3a_1}+1)\nonumber \\
    &+& \frac{3a_0a_1b_1}{32\pi b_2}-\frac{\omega a_0b_1}{16\pi a_1b_2}(1-\frac{3}{2}a_1)^2+\frac{a_0b_1}{8\pi a_1b_2}(1-\frac{3}{2}a_1)
  \end{eqnarray}
  which is automatically satisfied; and
  \begin{equation}\label{31}
    \alpha=-(1+\frac{2}{3a_1})
  \end{equation}
  In this case

  \begin{equation}\label{32}
    \rho=\frac{2b_3}{3a_1z^2}(b_0+b_1t)^{-1-\frac{3}{2}a_1}
  \end{equation}
  And the interaction $Q$ is given by
  \begin{equation}\label{33}
    Q=\frac{3z^2a_0b_1^2}{16b_2\pi}\bigg[\frac{4}{3}a_1^2-\frac{\omega}{2}(1-\frac{3}{2}a_1)^2+\frac{3}{2}a_1(1-\frac{3}{2}a_1)\bigg]
    (b_0+b_1t)^{-2-\frac{3}{2}a_1}
  \end{equation}
The physical and kinematical properties of the model are obtained as follows:\\ \\
Volume
\begin{equation}\label{34}
  V=b_2(b_0b_1t)^{\frac{3}{2}a_1}
\end{equation}
Hubble's parameter
\begin{equation}\label{35}
  H=\frac{1}{2}a_1b_1(b_0+b_1t)^{-1}
\end{equation}
Expansion factor
\begin{equation}\label{36}
  \theta=\frac{3}{2}a_1b_1(b_0+b_1t)^{-1}
\end{equation}
Deceleration parameter
\begin{equation}\label{37}
  q=\frac{2}{a_1}-1
\end{equation}
Jerk parameter
\begin{equation}\label{38}
  j=\frac{2b_1}{a_1}\left(\frac{a_1}{2}-1\right)\left(\frac{a_1}{2}-2\right)(b_0+b_1t)^{-1}
\end{equation}
And state-finder parameters $\{r, s\}$ are obtained as
\begin{equation}\label{39}
  r=4a_1^2\left(\frac{a_1}{2}-1\right)\left(\frac{a_1}{2}-2\right)
\end{equation}
\begin{equation}\label{40}
  s=\frac{2}{3}a_1^{-1}(a_1-2)(a_1-4)(4-3a_1)^{-1}
\end{equation}
Dark energy parameter
\begin{eqnarray}\label{41}
  \Omega_d &=& \frac{\rho_d}{3H^2}\nonumber \\
   &=& a_1^{-2}b_1^{-2}b_3(4+3a_1)(b_0+b_1t)^{-1-\frac{3}{2}a_1}+\frac{4}{3}a_1^{-2}b_1^{-2}b_4(b_0+b_1t)^3
\end{eqnarray}
  \subsection{Case-II:}
  As another solution we get from \eqref{17} and \eqref{18},
  \begin{equation}\label{42}
    \lambda=(c_1t+c_0)^{c_2}
  \end{equation}

  \begin{equation}\label{43}
    \phi=\frac{a_0}{c_1c_2}(c_1t+c_0)^{1-c_2}e^{\frac{-3}{2}(c_1t+c_0)^{c_2}}
  \end{equation}
  where $c_0$, $c_1$ and $c_2$ are arbitrary constants. Here in the case we take the interaction $Q$ as
  \begin{equation}\label{44}
    Q=3z^2H\rho_d
  \end{equation}
  where $H$ is the mean Hubble's parameter. Then equations \eqref{15} and \eqref{16} respectively take the forms
  \begin{equation}\label{45}
    \dot{\rho_m}+(\rho_m+p_m)\frac{3}{2}\dot{\lambda}=-3z^2\rho_d\frac{\dot{\lambda}}{2}
  \end{equation}
  and
  \begin{equation}\label{46}
    \dot{\rho_d}+(\rho_d+p_d)\frac{3}{2}\dot{\lambda}=3z^2\rho_d\frac{\dot{\lambda}}{2}
  \end{equation}
From \eqref{46} we get
\begin{equation}\label{47}
  \rho_d=c_3e^{\frac{3}{2}(z^2-\alpha-1)(c_1t+c_0)^{c_2}}
\end{equation}
where $c_3$ is an arbitrary constant.
Now using \eqref{42}, \eqref{43}, \eqref{47} in equation \eqref{9} we get
\begin{eqnarray} \label{48}
  \rho_m &=& \frac{a_0c_1c_2}{8\pi}\left(\frac{3}{4}-\frac{9\omega}{8}\right)(c_1t+c_0)^{c_2-1}e^{-\frac{3}{2}(c_1t+c_0)^{c_2}} \nonumber \\
   &-& \frac{\omega a_0c_1}{16\pi c_2}(1-c_2)^2(c_1t+c_0)^{-1-c_2}e^{-\frac{3}{2}(c_1t+c_0)^{c_2}} \nonumber \\
  &+& \frac{3a_0c_1}{16\pi}(1-c_2)(1+\omega)(c_1t+c_0)^{-1}e^{-\frac{3}{2}(c_1t+c_0)^{c_2}} \nonumber \\
   &-& \frac{9a_0c_1c_2}{32\pi}(c_1t+c_0)^{c_2-1}e^{-\frac{3}{2}(c_1t+c_0)^{c_2}}\nonumber \\
   &-& c_3e^{\frac{3}{2}(z^2-\alpha-1)(c_1t+c_0)^{c_2}}
\end{eqnarray}
Here \eqref{12} gives, using \eqref{47}, \eqref{48} and \eqref{13},
\begin{eqnarray} \label{49}
  p_m &=& \left(\frac{21a_0c_1c_2}{96\pi}+\frac{27\omega a_0c_1c_2}{48\pi}-\frac{9}{32\pi}-\frac{9\omega}{48\pi}\right)\nonumber \\
   &\times & (c_1t+c_0)^{c_2-1}e^{-\frac{3}{2}(c_1t+c_0)^{c_2}}+\frac{a_0c_1}{16\pi} (1-c_2)(1+\omega)(c_1t+c_0)^{-1}e^{-\frac{3}{2}(c_1t+c_0)^{c_2}}
   \nonumber \\
   &+& \bigg[\frac{a_0c_1^2(1-c_2)}{24\pi}(3+2\omega)-\frac{\omega a_0c_1(1-c_2)^2}{48\pi c_2}\bigg](
   c_1t+c_0)^{-c_2-1}e^{-\frac{3}{2}(c_1t+c_0)^{c_2}} \nonumber \\
  &-& \alpha c_3e^{\frac{3}{2}(z^2-\alpha-1)(c_1t+c_0)^{c_2}}
\end{eqnarray}
In this case
\begin{eqnarray}\label{50}
  \rho &=& \frac{a_0c_1c_2}{8\pi}\left(\frac{3}{4}-\frac{9\omega}{8}\right)(c_1t+c_0)^{c_2-1}e^{-\frac{3}{2}(c_1t+c_0)^{c_2}}\nonumber \\
   &-& \frac{\omega a_0c_1}{16\pi c_2}(1-c_2)^2(c_1t+c_0)^{-1-c_2}e^{-\frac{3}{2}(c_1t+c_0)^{c_2}}\nonumber \\
   &+& \frac{3a_0c_1}{16\pi}(1-c_2)(1+\omega)(c_1t+c_0)^{-1}e^{-\frac{3}{2}(c_1t+c_0)^{c_2}} \nonumber\\
   &-& \frac{9a_0c_1c_2}{32\pi}(c_1t+c_0)^{c_2-1}e^{-\frac{3}{2}(c_1t+c_0)^{c_2}}
\end{eqnarray}
And the interaction $Q$ is obtained as
\begin{equation}\label{51}
  Q=\frac{3}{2}z^2c_1c_2c_3(c_1t+c_0)^{c_2-1}e^{\frac{3}{2}(z^2-\alpha-1)(c_1t+c_0)^{c_2}}
\end{equation}
\subsection{Case-II(a):}
In this case we take the interaction $Q$ as $3z^2\rho_m\frac{\dot{\lambda}}{2}$, so that the conservation equations take the forms
\begin{equation}\label{52}
  \dot{\rho_m}+(\rho_m+p_m)\frac{3}{2}\dot{\lambda}=-3z^2\rho_m\frac{\dot{\lambda}}{2}
\end{equation}

\begin{equation}\label{53}
  \dot{\rho_d}+(\rho_d+p_d)\frac{3}{2}\dot{\lambda}=3z^2\rho_m\frac{\dot{\lambda}}{2}
\end{equation}
Now from \eqref{9} we get, using \eqref{42} and \eqref{43},
\begin{eqnarray}\label{54}
  \rho_m &=& \frac{a_0}{8\pi c_1c_2}e^{-\frac{3}{2}(c_1t+c_0)^{c_2}}\bigg[\left(\frac{3}{2}\omega c_1^2c_2(1-c_2)+\frac{3}{2}c_1^1c_2(1-c_2)\right)
   \nonumber \\
   &\times & (c_1t+c_0)^{-1}-\frac{\omega}{2}c_1^2(1-c_2)^2(c_1t+c_0)^{-1-c_2}-\left(\frac{3}{2}+\frac{9\omega}{8}\right)c_1^2c_2^2(c_1t+c_0)^{c_2-1}\bigg]
   -\rho_d
\end{eqnarray}
From \eqref{53} we have, using \eqref{54}, \eqref{13} and \eqref{42},
\begin{eqnarray}\label{55}
  \dot{\rho_d} &=& -\frac{3}{2}c_1c_2(c_1t+c_0)^{c_2-1}\rho_d+\frac{3a_0z^2}{16\pi}e^{-\frac{3}{2}(c_1t+c_0)^{c_2}}\bigg[
  -\frac{\omega}{2}c_1^2(1-c_2)^2(c_1t+c_0)^{-2}\nonumber \\
  &+& \frac{3}{2}c_1^2c_2(1+\omega)(1-c_2)(c_1t+c_0)^{c_2-2}-\left(\frac{3}{2}+\frac{9\omega}{8}\right)c_1^2c_2^2(c_1t+c_0)^{2c_2-1}\bigg],
\end{eqnarray}
where
\begin{equation}\label{56}
  \alpha=-z^2
\end{equation}
And this is possible without loss of generality as $\alpha$ can take values such that $-1\le \alpha<0$ as well as $\alpha<-1$ which is the characteristic
of different forms of dark energy which can be attained according to different values of $z^2$. Now equation \eqref{55} gives
\begin{eqnarray}\label{57}
  \rho_d &=& \frac{3a_0z^2}{16\pi}e^{-\frac{3}{2}(c_1t+c_0)^{c_2}}\bigg[\frac{\omega}{2}c_1(1-c_2)^2(c_1t+c_0)^{-1}\nonumber \\
   &-& \frac{1}{2}\left(\frac{3}{2}+\frac{9\omega}{8}\right)c_1c_2(c_1t+c_0)^{2c_2}-\frac{3}{2}(1+\omega)c_1c_2(c_1t+c_0)^{c_2-1}\bigg]
\end{eqnarray}
 From \eqref{54} and \eqref{57} we have
 \begin{eqnarray}\label{58}
   \rho_m &=& e^{-\frac{3}{2}(c_1t+c_0)^{c_2}}\bigg[\left(\frac{3a_0c_1}{16\pi}(1+\omega)(1-c_2)-\frac{3a_0\omega c_1z^2}{32\pi}(1-c_2)^2\right)\nonumber \\
    &\times & (c_1t+c_0)^{-1}+\left(\frac{9a_0c_1c_2}{32\pi}(1+\omega)z^2-\frac{a_0}{8\pi}c_1c_2\left(\frac{3}{2}+\frac{9\omega}{8}\right)\right)
    (c_1t+c_0)^{c_2-1}\nonumber \\
    &+& \frac{3a_0c_1c_2}{32\pi}z^2\left(\frac{3}{2}+\frac{9\omega}{8}\right)(c_1t+c_0)^{2c_2}-\frac{a_0\omega c_1}{16\pi c_2}(1-c_2)^2(c_1t+c_0)^{-c_2-1}\bigg]
 \end{eqnarray}
 Now from \eqref{57} and \eqref{13} we take
\begin{eqnarray} \label{59}
  p_d &=& \frac{3a_0z^2\alpha}{16\pi}e^{-\frac{3}{2}(c_1t+c_0)^{c_2}}\bigg[\frac{\omega}{2}c_1(1-c_2)^2(c_1t+c_0)^{-1}\nonumber \\
   &-& \frac{1}{2}\left(\frac{3}{2}+\frac{9\omega}{8}\right)c_1c_2(c_1t+c_0)^{2c_2}-\frac{3}{2}(1+\omega)c_1c_2(c_1t+c_0)^{c_2-1}\bigg]
\end{eqnarray}
Thus now using \eqref{59} in equation \eqref{10} we get
\begin{eqnarray}\label{60}
  p_m &=& \bigg[\{\frac{3a_0\omega c_1}{32\pi}z^4(1-c_2)^2+\frac{3\omega a_0c_1}{16\pi}(1-c_2)\nonumber \\
   &+& \frac{3a_0c_1}{16\pi}(1-c_2)\}(c_1t+c_0)^{-1}-\frac{3a_0z^4c_1c_2}{32\pi}\left(\frac{3}{2}+\frac{9\omega}{8}\right)(c_1t+c_0)^{2c_2}\nonumber \\
  &+& \left(\frac{9a_0z^4}{32\pi}(1+\omega)c_1c_2+\frac{3a_0c_1c_2}{32\pi}-\frac{9\omega a_0c_1c_2}{32\pi}-\frac{9a_0c_1}{32\pi}\right)
  (c_1t+c_0)^{c_2-1}\nonumber \\
   &+& \left(\frac{a_0c_1(1-c_2)}{8\pi}-\frac{\omega a_0c_1}{16\pi c_2}(1-c_2)^2\right)(c_1t+c_0)^{-1-c_2}\bigg]e^{-\frac{3}{2}(c_1t+c_0)^{c_2}}
\end{eqnarray}
And in this case
\begin{eqnarray}\label{61}
  \rho &=& \frac{a_0}{8\pi c_1c_2}e^{-\frac{3}{2}(c_1t+c_0)^{c_2}}\bigg[\left(\frac{3\omega}{2}c_1^2c_2(1-c_2)+\frac{3}{2}c_1^2c_2(1-c_2)\right)
  \nonumber \\
   &\times & (c_1t+c_0)^{-1}-\frac{\omega}{2}c_1^2(1-c_2)^2(c_1t+c_0)^{-1-c_2}-\left(\frac{3}{2}+\frac{9\omega}{8}\right)c_1^1c_2^2(c_1t+c_0)^{c_2-1}\bigg]
\end{eqnarray}
Here the interaction $Q$ is obtained as
\begin{eqnarray}\label{62}
  Q &=& \frac{3}{2}c_1c_2z^2(c_1t+c_0)^{c_2-1}e^{-\frac{3}{2}(c_1t+c_0)^{c_2}}\nonumber \\
   &\times & \bigg[\left(\frac{3a_0c_1}{16\pi}(1+\omega)(1-c_2)-\frac{3a_0\omega c_1z^2(1-c_2)^2}{32\pi}\right)(c_1t+c_0)^{-1} \nonumber \\
   &+& \left(\frac{9a_0c_1c_2}{32 \pi}(1+\omega)z^2-\frac{a_0}{8\pi}c_1c_2\left(\frac{3}{2}+\frac{9\omega}{8}\right)\right)(c_1t+c_0)^{c_2-1}
   \nonumber \\
   &+& \frac{3a_0c_1c_2}{32\pi}z^2\left(\frac{3}{2}+\frac{9\omega}{8}\right)(c_1t+c_0)^{2c_2}-\frac{a_0\omega c_1}{16\pi c_2}(1-c_2)^2(c_1t+c_0)^{-1-c_2}
   \bigg]
\end{eqnarray}
The physical and kinematical properties of the models are as follows
\begin{equation}\label{63}
  V=e^{\frac{3}{2}(c_1t+c_0)^{c_2}}
\end{equation}
\begin{equation}\label{64}
  H=\frac{1}{2}c_1c_2(c_1t+c_0)^{c_2-1}
\end{equation}
\begin{equation}\label{65}
  \theta=\frac{3}{2}c_1c_2(c_1t+c_0)^{c_2-1}
\end{equation}
\begin{equation}\label{66}
  q=-\frac{2}{c_1c_2}(c_2-1)(c_1t+c_0)^{-c_2}-1
\end{equation}
\begin{equation}\label{67}
  j=\frac{1}{2}c_1c_2(c_1t+c_0)^{c_2-1}+3c_1(c_2-1)(c_1t+c_0)^{-1}+2c_1c_2^{-1}(c_1t+c_0)^{-c_2-1}
\end{equation}

\begin{equation}\label{68}
 r= 1+4c_2^{-2}(c_2-1)(c_2-2)(c_1t+c_0)^{-2c_2}+6c_2^{-1}(c_2-1)(c_1t+c_0)^{-c_2}
\end{equation}
\begin{eqnarray}\label{69}
  s &=& -\frac{1}{3}\bigg[\frac{2}{c_1c_2}(c_2-1)(c_1t+c_0)^{-c_2}+\frac{3}{2}\bigg]^{-1}\nonumber \\
   &\times& \bigg[6c_2^{-1}(c_2-1)(c_1t+c_0)^{-c_2}+4c_2^{-2}(c_2-1)(c_2-2)(c_1t+c_0)^{-2c_2}\bigg]
\end{eqnarray}
\subsection{Case-III}
Equations \eqref{17} and \eqref{18} give
\begin{equation}\label{70}
  \lambda=\beta (\log t)^n, ~~\beta>0,~n>1
\end{equation}

\begin{equation}\label{71}
  \phi=a_0(n\beta)^{-1}t(\log t)^{1-n}e^{-\frac{3\beta}{2}(\log t)^n}
\end{equation}
Here in this case we assume the interaction between dark energy and other matters of the universe in the form
\begin{equation}\label{72}
  Q=3z^2H\rho_m
\end{equation}
so that equations \eqref{15} and \eqref{16} respectively take the forms

\begin{equation}\label{73}
  \dot{\rho_m}+(\rho_m+p_m)3\frac{\dot{\lambda}}{2}=-3z^2\rho_m\frac{\dot{\lambda}}{2}
\end{equation}

\begin{equation}\label{74}
  \dot{\rho_d}+(\rho_d+p_d)3\frac{\dot{\lambda}}{2}=3z^2\rho_m\frac{\dot{\lambda}}{2}
\end{equation}
Now from equation \eqref{9} we get
\begin{eqnarray}\label{75}
  \rho_m &=& \frac{a_0n\beta}{8\pi t}(\log t)^{n-1}e^{-\frac{3}{2}\beta (\log t)^n}+\frac{3a_0}{16\pi t}(1-\omega)(1-n)(\log t)^{-1}
  e^{-\frac{3}{2}\beta (\log t)^n}\nonumber \\
  &+& \frac{a_0}{8\pi t}\left(1-\frac{\omega}{2}\right)e^{-\frac{3}{2}\beta (\log t)^n}-\frac{a_0\omega}{16\pi n\beta t}(\log t)^{1-n}
  e^{-\frac{3}{2}\beta (\log t)^n}\nonumber \\
   &-& \frac{(1-n)^2\omega a_0}{16\pi n\beta t}(\log t)^{-n-1}e^{-\frac{3}{2}\beta (\log t)^n} -
   \frac{(1-n)\omega a_0}{8\pi n\beta t}(\log t)^{-n}e^{-\frac{3}{2}\beta (\log t)^n}-\rho_d
\end{eqnarray}
Thus from equations \eqref{74} and \eqref{75}, using relation \eqref{13}, we have
\begin{equation}\label{76}
  \rho_d=e^{-\frac{3}{2}\beta(\alpha+1+z^2)(\log t)^n}\times \psi (t)
\end{equation}
where
\begin{eqnarray}\label{77}
  \psi (t) &=& \int \frac{3z^2n\beta}{2t^2}(\log t)^{n-1}e^{\frac{3}{2}\beta(\alpha+z^2)(\log t)^n}
    \bigg[\frac{na_0\beta}{8\pi}(\log t)^{n-1}+\frac{3a_0(1-n)}{16\pi}(1-\omega)(\log t)^{-1} \nonumber\\
   &+& \left(1-\frac{\omega}{2}\right)\frac{a_0}{8\pi}-\frac{(1-n)^2\omega a_0}{16\pi n\beta}(\log t)^{-n-1} \nonumber\\
   &-&\frac{\omega a_0}{16\pi n\beta}(\log t)^{1-n}-\frac{(1-n)\omega a_0}{8\pi n\beta}(\log t)^{-n}\bigg]dt
\end{eqnarray}
Again using relations \eqref{70}, \eqref{71}, \eqref{75}, \eqref{76} and \eqref{13} in equation \eqref{12} we have
\begin{eqnarray}\label{78}
  p_m &=& \frac{1}{24\pi t}(3+2\omega)e^{-\frac{3}{2}\beta(\log t)^{n}}\bigg[\frac{a_0}{\beta}(1-n)(\log t)^{-1-n}+\frac{3}{2}a_0+\frac{3}{2}a_0(1-n)
  (\log t)^{-1}\nonumber \\
   &-& \frac{a_0}{n \beta}(1-n)(\log t)^{-n}-\frac{9na_0\beta}{4}(\log t)^{n-1}\bigg]-\frac{n\beta(3+2\omega)}{16\pi t}(\log t)^{n-1}
   e^{-\frac{3}{2}\beta(\log t)^{n}} \nonumber\\
   &\times & \bigg[\frac{a_0}{n\beta}(\log t)^{1-n}-\frac{3a_0}{2}+\frac{a_0}{n\beta}(1-n)(\log t)^{-n}\bigg]+\frac{1}{3}\bigg[
   \frac{na_0\beta}{8\pi t}(\log t)^{n-1}\nonumber \\
   &+& \frac{3(1-\omega)a_0(1-n)}{16\pi t}(\log t)^{-1}+\frac{a_0}{8\pi t}(1-\frac{\omega}{2})-\frac{\omega a_0}{16 \pi n\beta t}(\log t)^{1-n}
   -\frac{(1-n)^2\omega a_0}{16\pi n \beta t}(\log t)^{-1-n}\nonumber \\
   &-& \frac{(1-n)\omega a_0}{8 \pi n\beta t}(\log t)^{-n}\bigg]e^{-\frac{3}{2}\beta(\log t)^n}-\alpha \psi(t)
   e^{-\frac{3}{2}\beta(\alpha+1+z^2)(\log t)^n}
\end{eqnarray}
Here
\begin{eqnarray}\label{79}
  \rho_m &=& \frac{na_0\beta}{8\pi t}(\log t)^{n-1}e^{-\frac{3}{2}\beta(\log t)^n}+\frac{3a_0}{16\pi t}(1-\omega)(1-n)(\log t)^{-1}
  e^{-\frac{3}{2}\beta(\log t)^n}\nonumber \\
   &+& \frac{a_0}{8\pi t}(1-\frac{\omega}{2})e^{-\frac{3}{2}\beta(\log t)^n}-\frac{\omega a_0}{16\pi n\beta t}(\log t)^{1-n}
   e^{-\frac{3}{2}\beta(\log t)^n}\nonumber \\
   &-& \frac{\omega a_0}{16\pi n\beta t}(1-n)^2(\log t)^{-n-1}e^{-\frac{3}{2}\beta(\log t)^n}
   -\frac{\omega a_0(1-n)}{8\pi n\beta t}(\log t)^{-n}e^{-\frac{3}{2}\beta(\log t)^n}\nonumber \\
   &-& e^{-\frac{3}{2}\beta(\alpha +1+z^2)(\log t)^n}\psi(t)
\end{eqnarray}
And
\begin{eqnarray}\label{80}
  \rho &=& \frac{a_0n\beta}{8\pi t}(\log t)^{n-1}e^{-\frac{3}{2}\beta(\log t)^n}+\frac{3a_0}{16\pi t}(1-\omega)(1-n)(\log t)^{-1}
  e^{-\frac{3}{2}\beta(\log t)^n}\nonumber \\
   &+& \frac{a_0}{8\pi t}(1-\frac{\omega}{2})e^{-\frac{3}{2}\beta(\log t)^n}-\frac{\omega a_0}{16\pi n\beta t}(\log t)^{1-n}
   e^{-\frac{3}{2}\beta(\log t)^n}\nonumber \\
   &-& \frac{(1-n)^2\omega a_0}{16\pi n \beta t}(\log t)^{-n-1}e^{-\frac{3}{2}\beta(\log t)^n}-\frac{(1-n)\omega a_0}{8\pi n\beta t}
   (\log t)^{-n}e^{-\frac{3}{2}\beta(\log t)^n}
\end{eqnarray}
In this case the interaction $Q$ is given by
\begin{eqnarray}\label{81}
  Q &=& \frac{3z^2n\beta}{2t}(\log t)^{n-1}\bigg[\frac{na_0\beta}{8\pi t}(\log t)^{n-1}e^{-\frac{3}{2}\beta(\log t)^{n}}+
  \frac{3a_0}{16\pi t}(1-\omega)(1-n)(\log t)^{-1}e^{-\frac{3}{2}\beta(\log t)^{n}}\nonumber \\
   &+& \frac{a_0}{8\pi t}(1-\frac{\omega}{2})e^{-\frac{3}{2}\beta(\log t)^{n}} -\frac{\omega a_0}{16\pi n\beta t}
   (\log t)^{1-n}e^{-\frac{3}{2}\beta(\log t)^{n}}\nonumber\\
   &-& \frac{\omega a_0}{16\pi n\beta t}(1-n)^2(\log t)^{-n-1}e^{-\frac{3}{2}\beta(\log t)^{n}}-
   \frac{\omega a_0(1-n)}{8\pi n\beta t}(\log t)^{-n}e^{-\frac{3}{2}\beta(\log t)^{n}}\nonumber \\
   &-& e^{-\frac{3}{2}\beta(\alpha+1+z^2)(\log t)^{n}}\psi(t)\bigg]
\end{eqnarray}
The physical and kinematical properties of the model are obtained as follows

\begin{equation}\label{82}
  V=e^{\frac{3}{2}\beta (\log t)^{n}}
\end{equation}

\begin{equation}\label{83}
  H=\frac{n\beta}{2t}(\log t)^{n-1}
\end{equation}

\begin{equation}\label{84}
  \theta=\frac{3n\beta}{2t}(\log t)^{n-1}
\end{equation}
\begin{equation}\label{85}
  q=\frac{2}{n\beta}(\log t)^{n-1}+\frac{2(n-1)}{n\beta}(\log t)^{n-2}-1
\end{equation}
\begin{eqnarray}\label{86}
  j &=& \frac{n\beta}{2}t^{-1}(\log t)^{n-1}+3(n-1)t^{-3}(\log t)^{-1}+\frac{4}{n\beta}t^{-1}(\log t)^{-n+1}\nonumber \\
   &-& 3t^{-3}-\frac{6}{n\beta}(n-1)t^{-1}(\log t)^{-n}+\frac{2}{n\beta}(n-1)(n-2)t^{-1}(\log t)^{n-3}
\end{eqnarray}

\begin{eqnarray}\label{87}
  r &=& 1+\frac{6}{n\beta}(n-1)t^{-2}(\log t)^{-n}-\frac{6}{n\beta}t^{-2}(\log t)^{-n+1}+\frac{8}{n^2\beta^2}(\log t)^{-2n+2} \nonumber\\
   &-& \frac{12}{n^2\beta^2}(n-1)(\log t)^{-2n+1}+\frac{4}{n^2\beta^2}(n-1)(n-2)(\log t)^{-2}
\end{eqnarray}
\begin{eqnarray}\label{88}
  s &=& \frac{1}{3}\bigg[\frac{2}{n\beta}(\log t)^{n-1}+\frac{2(n-1)}{n\beta}(\log t)^{n-2}-\frac{3}{2}\bigg]^{-1}\nonumber \\
   &\times & \bigg[\frac{6}{n\beta}(n-1)t^{-2}(\log t)^{-n}-\frac{6}{n\beta}t^{-2}(\log t)^{-n+1}+\frac{8}{n^2\beta^2}(\log t)^{-2n+2}\nonumber \\
   &-& \frac{12}{n^2\beta^2}(n-1)(\log t)^{-2n +1}+\frac{4}{n^2\beta^2}(n-1)(n-2)(\log t)^{-2}
\end{eqnarray}
\section{Study of the solutions and conclusions}
For the model universe in Case-I we see that, at $t=0$, the energy density has finite value dependent on the coupling constant of the interaction between
dark energy and other matters in this universe, and the total energy density is found to decrease with time until it tends to zero at infinite time and the
interaction term is also found to follow the same behavior with respect to time. But during this time the rate of decrease of the dark energy density is slower in comparison to the rate of decrease of the energy density of other matters in the universe. Therefore as time passes by, dark energy seems to
dominate over other matters. Thus the scenario in such type of universe is that dark energy plays a vital role and it seems that as the
energy density of dark energy increases the expansion of the universe increases.

In this model it is seen that, at $t=0$, $V=b_2(b_0+b_1t)^{\frac{3}{2}a_1}$, which shows that, if we have to accept the big bang theory, the
universe begins its evolution at time $t\to \frac{-b_0}{b_1}$, thereby implying the existence of negative time which is almost possible from the presence
of dark energy in this universe. Again the value of the deceleration parameter obtained here implies that the value of $a_1$ is limited by the condition
$(a_1>2)$. And in this model the expansion of the universe is accelerating though the rate decreases with time. Therefore this universe may be taken
as a reasonable model; otherwise if the rate of expansion increases with time there will be a singularity where the universe ends transforming itself into a
 cloud of dust.

 In this universe we see that the interaction between dark energy and other matters decreases with time, and perhaps there is a tendency where the
 dark energy decays into cold dark matter. Again it is seen that dark energy density is zero only at time $t=\frac{-b_0}{b_1}$ showing that dark energy
 exists before $t=0$ also. Thus perhaps there exist a epoch before our cosmic time begins in the history of evolution of the universe.

 Here in this case, if either $a_1=2$ or $a_1=4$, then we get the state-finder parameter $\{r, s\}$ as $r=0$ and $s=0$ for which the dark energy
 model reduces to a flat $\Lambda CDM$ model which predicts a highly accelerated expansion before these events of time. In this model the interaction between dark energy and other matters are found to exist at these events of time not interrupted by the high speed of expansion.

 For this universe the equation of state parameter for dark energy is found to be less than $-1$ which indicates that the dark energy contained is
 of the phantom type. From the study of the interaction we also
 see that the action of such type of dark energy is more when the energies from other types of sources remain idle or not so active. It is also
 opposite to or against the light energy. It acts also against the living energy or the energy possessed by the human beings.

 In Case-II, though the universe is expanding and the rate of expansion is accelerating it depends much on the value of $c_1$, which indicates that the
 expansion is related to the dark energy density. There it may be taken that dark energy enhances the accelerated expansion. And this enhancement is also
 dependent on the value of $z^2$ which is the coupling constant of the interaction between dark energy and other matters in this universe. This implies
 that the expansion of the universe is very much interrelated with the interaction between dark energy and other components of the universe. Thus all the
 members of this universe may be taken to expand due to also the presence of dark energy. Hence, considering on a small scale structure of the universe,
 the earth may be taken to expand due to also the presence of dark energy.

 In this universe we see that, when $c_2\to 0$ it goes to the asymptotic static era with $r\to \infty$ and $s\to\-\infty$. And when $c_2=1$, the universe
 goes to $\Lambda CDM$ model for which $r=1$ and $s=0$. Thus the state-finder parameters $\{r, s\}$ show the picture of the evolution of our universe,
 starting from the asymptotic static era and then coming to the $\Lambda CDM$ model era. Here we see that the interaction $Q\to 0$ as $c_2\to 0$ which
 means that the interaction almost stops at the cosmic time when the scale factor of the universe becomes or takes the value $e^{\frac{1}{2}}$.

 Again it is seen that the energy density of this model universe tends to infinity at $c_2=0$, and decreases gradually as $c_2\to 1$, thus it seems that
 our universe started with a big bang. From the above behaviors of this model it is also implied that at the beginning of the evolution of this universe
 there was no interaction between dark energy and other components of the universe, and after that the interaction between them becomes active and increases
 with time, and at present they are highly interacting. But this interaction also depends much on the values of $\alpha$ and $z^2$ which are
 respectively equation of state parameter of the dark energy contained and the coupling constant of interaction.

 Case-III represents the logamediate scenario of the universe where the cosmological solutions have indefinite expansion \cite{Bilic}. In this case
 the dark energy has a quintessence like behavior. Here the matter content of the universe is seen to increase slowly due to the interaction and the
 cosmic effect. In this universe there is an interesting event of time, that is, the cosmic time when $t=1$. At this point the universe has
 the tendency of accelerating expansion, but the expansion suddenly stops at this moment and the volume of the universe takes the value 1 at this
 juncture. So it seems that there is a bounce and a new epoch begins from this juncture. Also we see that, at $t=1$, the state-finder parameters
 $\{r, s\}$ take the values $r=1$, $s=0$. Thus at this instant our universe will go to that of a $\Lambda CDM$ model which implies that at this event
 of time most of the dark energy contained will reduce to cold dark matter.

 The energy density of this model universe tends to infinity at $t=0$ which indicates that this universe begins with a big bang, and it (energy density)
 decreases gradually until it tends to a finite quantity at infinite time, of course with a bounce at $t=1$. And at $t=0$, the dark energy density
 is found to be exceptionally high which indicates that it helps much in triggering the big bang. It is also seen that the dark energy is
 highly interacting with other components of the universe at $t=0$, the interaction decreasing slowly with the passing away of the cosmic time.
 In this universe we see that both the scalar field $\phi$ and the interaction $Q$ tend to vanish as $a_0\to 0$. Thus the scalar field is very
 much interconnected with the dark energy content of this universe and plays a vital role in the production and existence of it. One peculiarity
 in this model is that the scalar field does not vanish at $t\to \infty$; thus the dark energy seems to be prevalent eternally in this universe
 due to this scalar field.


\begin{thebibliography}{}

\bibitem{Riess} A. G. Riess et al., Astron. J. \textbf{116}, 1009 (1998).

\bibitem{Perlmutter} S. Perlmutter et al. Astrophys. J. \textbf{517}, 565 (1999).

\bibitem{Amendola} L. Amendola and S. Tsujikawa, Dark Energy: Theory and Observations, Cambridge University Press, Cambridge UK (2010).

\bibitem{Weinberg} S. Weinberg, Rev. Mod. Phys. \textbf{61}, 1 (1989).

\bibitem{Brans} C. Brans and R. H. Dicke, Phys. Rev. \textbf{124}, 925 (1961).

\bibitem{Saez} D. Saez and V. J. Balester, Phys. Lett. A \textbf{113}, 467 (1986).

\bibitem{Barraco} D. E. Barraco and V. H. Hamity, Phys. Rev. D \textbf{62}, 044027 (2000).

\bibitem{Capozziello} S. Capozziello, Int. J. Mod. Phys. D \textbf{11}, 483 (2002).

\bibitem{Capozziello1} S. Capozziello, Vincenzo F. Cardone and A. Troisi,  Phys. Rev. D \textbf{71}, 043503 (2005).

\bibitem{Capozziello2}  S. Capozziello, Vincenzo F. Cardone and M. Francaviglia, Gen. Rel. Grav. \textbf{38}, 711 (2006).

\bibitem{Nojiri} S. Nojiri and Sergei D. Odintsov,  Phys. Rev. D \textbf{74}, 086005 (2006).   	

\bibitem{Bamba} Kazuharu Bamba and Chao-Qiang Geng, Phys.Lett. B \textbf{679}, 282 (2009).	

\bibitem{Ferraro} R. Ferraro and F. Fiorini, Phys. Rev. D \textbf{75}, 084031 (2007).

\bibitem{Myrzakulov}R. Myrzakulov, Eur. Phys. J. C \textbf{71}, 1752 (2011).

\bibitem{Bamba1} Kazuharu Bamba, Chao-Qiang Geng, Chung-Chi Lee and Ling-Wei Luo, JCAP \textbf{1101}, 021 (2011).

\bibitem{Setare}M. R. Setare and F. Darabi,  Gen. Rel. Grav. \textbf{44}, 2521 (2012).  	
\bibitem{Nojiri1} S. Nojiri, S. D. Odintsov and M. Sasaki, Phys. Rev. D \textbf{71}, 123509 (2005).
  	
\bibitem{Sokolowski}L. M. Sokolowski, Z. A. Golda, M. Litterio and L. Amendola, Int. J. Mod. Phys. A \textbf{6}, 4517 (1991).
  	
\bibitem{Iglesias} A. Iglesias and Z. Kakushadze, Int. J. Mod. Phys. A \textbf{16}, 3603 (2001).
\bibitem{Nojiri2}S. Nojiri, S. D. Odintsov and S. Ogushi, Int. J. Mod. Phys. A \textbf{17}, 4809, (2002).  	
\bibitem{Colgain}E. O. Colgain and H. Yavartanoo, JHEP \textbf{0908},  021 (2009). 	
\bibitem{Ghodsi}A. Ghodsi, Int. J. Mod. Phys. A \textbf{26}, 925, (2011).
  	
\bibitem{Park}Mu-in Park, JCAP \textbf{1001}, 001 (2010).	
\bibitem{Harko}T. Harko, Francisco S. N. Lobo, S. Nojiri and Sergei D. Odintsov, Phys. Rev. D \textbf{84}, 024020 (2011). 	





 \bibitem{Dehghani} M. H. Dehghani, Phys. Rev. D \textbf{70}, 064009 (2004). 	

 \bibitem{Nojiri3} S. Nojiri and Sergei D. Odintsov, Gen. Rel. Grav. \textbf{36}, 1765 (2004).	


 \bibitem{Lue} A. Lue, R. Scoccimarro and G. Starkman, Phys. Rev. D \textbf{69}, 044005 (2004).

 	


\bibitem{Carroll} S. M. Carroll, Antonio De Felice, Vikram Duvvuri, Damien A. Easson, Mark Trodden and Michael S. Turner, Phys. Rev. D \textbf{71}, 063513 (2005).





\bibitem{Samanta}G. C. Samanta and S. N. Dhal, Int. J. theo. Phys. \textbf{52}, 1334 (2013).

\bibitem{Samanta1}G. C. Samanta, Int. J. Theo. Phys. \textbf{52}, 2647 (2013).

\bibitem{Samanta2}G. C. Samanta, Int. J. Theo. Phys. \textbf{52}, 2303 (2013).

\bibitem{Samanta3}G. C. Samanta, Int. J. Theo. Phys. \textbf{52}, 4389 (2013).

\bibitem{Singh} K. M. Singh and K. Priyokumar, Int. J. Theo. Phys. \textbf{53}, 4360 (2014).




\bibitem{Singh1} K. M. Singh and K. L. Mahanta, Astrophys. Space Sci. \textbf{361}, 85 (2016).








 \bibitem{Bamba3}K. Bamba, Int. J. Geom. Meth. Mod. Phys. \textbf{13}, 1630007 (2016). 	





 \bibitem{Zubair} M. Zubair, Syed M. Ali Hassan and G. Abbas, Can. J. Phys. \textbf{94}, 1289 (2016).	


 \bibitem{Maharaj}S. D. Maharaj, R. Goswami, S. V. Chervon and A. V. Nikolaev,  Mod. Phys. Lett. A \textbf{32}, 1750164 (2017). 	
  \bibitem{Priyokumar} K. Priyokumar, K. M. Singh and M. R. Mollah, Int. J. Theo. Phys. \textbf{56}, 2607 (2017).

 \bibitem{Samanta5}G. C. Samanta and R. Myrzakulov, Chin. J. Phys. \textbf{55}, 1044 (2017).	


 \bibitem{Morais} J. P. Morais Graça, Godonou I. Salako and Valdir B. Bezerra, Int. J. Mod. Phys. D \textbf{26}, 1750113 (2017).  	


 \bibitem{Bamba2}K. Bamba, Sergei D. Odintsov and Emmanuel N. Saridakis,  Mod. Phys. Lett. A \textbf{32}, 1750114 (2017). 	


 \bibitem{Noureen}I. Noureen and M. Zubair, Int. J. Mod. Phys. D \textbf{26}, 1750128 (2017). 	

  \bibitem{Samanta4}G. C. Samanta, R. Myrzakulov and P. Shah, Zeitschrift für Naturforschung A \textbf{72}, 365 (2017).

 \bibitem{Baffou} E. H. Baffou, M. J. S. Houndjo, M. Hamani-Daouda and F. G. Alvarenga, Eur. Phys. J. C \textbf{77}, 708 (2017).  	



 \bibitem{Azmat} Hina Azmat, M. Zubair and Ifra Noureen, Int. J. Mod. Phys. D \textbf{27}, 1750181 (2017). 	


 \bibitem{Singh2}J. K. Singh, Ritika Nagpal and S. K. J. Pacif, Int. J .Geom. Meth. Mod. Phys. \textbf{15}, 1850049 (2017). 	


 \bibitem{Odintsov} S. D. Odintsov and V. K. Oikonomou, Phys. Rev. D \textbf{96}, 104059 (2017). 	

 \bibitem{Addazi}A. Addazi, Int. J. Mod. Phys. A \textbf{33}, 1850030 (2018). 	

\bibitem{Shabani} H. Shabani and A. H. Ziaie, Int. J. Mod. Phys. A \textbf{33}, 1850050 (2018).	


 \bibitem{Odintsov1}S. D. Odintsov and V. K. Oikonomou, Annals Phys. \textbf{388}, 267 (2018).  	



 \bibitem{Zubair1} M. Zubair, Hina Azmat and Ifra Noureen, Int. J. Mod. Phys. D \textbf{27},  1850047 (2018).  	


 \bibitem{Kleidis} K. Kleidis and V. K. Oikonomou, Int. J. Geom. Meth. Mod. Phys. \textbf{15},  1850137 (2018). 	


 \bibitem{Tiwari} R. K. Tiwari, A. Beesham and B. Shukla. Int. J. Geom. Meth. Mod. Phys. \textbf{15}, 1850115 (2018)
 	

\bibitem{Hansraj}S. Hansraj and Ayan Banerjee, Phys. Rev. D \textbf{97}, 104020 (2018). 	



\bibitem{Rowland}D. Rowland and I. B. Whittingham, Mon. Not. R. Astron. Soc. \textbf{390}, 1719 (2008).
\bibitem{Cui} W. Cui et al, Mon. Not. R. Astron. Soc. \textbf{424}, 993 (2012).
\bibitem{Baldi} M. Baldi, Mon. Not. R. Astron. Soc. \textbf{414}, 116 (2011).
\bibitem{Sadeghi} J. Sadeghi et al, JCAP \textbf{12}, 031 (2013).
\bibitem{Khurshudyan} M. Khurshudyan et al, Int. J. Theor. Phys. \textbf{53}, 2370 (2014).

\bibitem{Bertolami}O. Bertolami, G. Pedro F. and Le Delliou M., Phys Lett. B, \textbf{654}, 165 (2007).
\bibitem{Olivares}G. Olivares, F. Atrio and D. Pavon, Phys. Rev. D \textbf{71}, 063523 (2005).

\bibitem{Bilic}
N. Bilic, G. B. Tupper and R. D. Viollier, Phys. Lett. B \textbf{535}, 17 (2002).


%
%
%
%
%
%
%






\end{thebibliography}
\end{document}